\documentclass[9pt,twocolumn,twoside]{revtex4-1}

\usepackage{textcomp}
\usepackage{tikz}
\usepackage{bm}
\usepackage{dcolumn}
\usepackage{booktabs}
\usepackage[colorlinks=true]{hyperref}
\usepackage[scaled]{helvet}
\usepackage{sansmath}
\usepackage{graphicx}
\usepackage{transparent}

\usepackage{amsmath}
\usepackage{amsfonts}
\usepackage{amssymb}
\usepackage{color}
\usepackage{gensymb}
\usepackage{multirow}
\usepackage{url}

\begin{document}

\title{Storage of hyperentanglement in a solid-state quantum memory}

\author{Alexey Tiranov$^{1}$}
\email{alexey.tiranov@unige.ch}
\author{Jonathan Lavoie$^{1}$}
\author{Alban Ferrier$^{2}$}
\author{Philippe Goldner$^{2}$}
\author{Varun B.~Verma$^{3}$}
\author{Sae Woo Nam$^{3}$}
\author{Richard P.~Mirin$^{3}$}
\author{Adriana E.~Lita$^{3}$}
\author{Francesco Marsili$^{4}$}
\author{Harald Herrmann$^{5}$}
\author{Christine Silberhorn$^{5}$}
\author{Nicolas Gisin$^{1}$}
\author{Mikael Afzelius$^{1}$}
\author{F\'{e}lix Bussi\`{e}res$^{1}$}

\affiliation{$^{1}$Group of Applied Physics, University of Geneva, CH-1211
  Geneva 4, Switzerland}
\affiliation{$^{2}$Chimie ParisTech, Laboratoire de Chimie de la Mati\`ere Condens\'ee de Paris, CNRS-UMR 7574, UPMC Univ Paris 06, 75005 Paris, France}
\affiliation{$^{3}$National Institute of Standards and Technology, 325 Broadway, Boulder, CO 80305, USA}
\affiliation{$^{4}$Jet Propulsion Laboratory, California Institute of Technology, 4800 Oak Grove Dr., Pasadena, California 91109, USA}
\affiliation{$^{5}$Applied Physics / Integrated Optics Group, University of Paderborn, 33095 Paderborn, Germany}


\date{\today}



\begin{abstract}
Two photons can simultaneously share entanglement between several degrees of freedom such as polarization, energy-time, spatial mode and orbital angular momentum. This resource is known as hyperentanglement, and it has been shown to be an important tool for optical quantum information processing. Here we demonstrate the quantum storage and retrieval of photonic hyperentanglement in a solid-state quantum memory.  
A pair of photons entangled in polarization and energy-time is generated such that one photon is stored in the quantum memory, while the other photon has a telecommunication wavelength suitable for transmission in optical fibre. 
We measured violations of a Clauser-Horne-Shimony-Holt (CHSH) Bell inequality for each degree of freedom, independently of the other one, which proves the successful storage and retrieval of the two bits of entanglement shared by the photons. Our scheme is compatible with long-distance quantum communication in optical fibre, and is in particular suitable for linear-optical entanglement purification for quantum repeaters.
\end{abstract}


\newcommand{\ket}[1]{\vert#1\rangle}
\newcommand{\bra}[1]{\langle#1\vert}
\newcommand{\YSO}{Y$_2$SiO$_5$}
\newcommand{\nd}[0]{Nd$^{3+}$:Y$_2$SiO$_5$}
\newcommand{\X}{\textbf{X}}
\newcommand{\Y}{\textbf{Y}}
\newcommand{\e}{\mathrm{e}}

\maketitle

\section{Introduction}
Quantum entanglement is an essential resource for quantum information processing, and in particular for quantum communication and for quantum computing. There are many ways in which quantum systems can be entangled. For example, two photons can be entangled in their polarization, or in their energy. They can also be entangled in more than one of their degrees of freedom (DOF), i.e.~hyperentangled~\cite{Kwiat1997,Barreiro2005,Barbieri2005a}. Two photons can thus share more entanglement bits (ebits) than what a singly-entangled pair allows. 

Hyperentanglement is an important resource in optical quantum information processing~\cite{Pan2012a}. For example, complete and deterministic Bell-state analysis in one of the DOF of a hyperentangled pair is possible with linear optics~\cite{Kwiat1998,Walborn2003a,Schuck2006}. This was used to perform quantum teleportation~\cite{Boschi1998a,Kim2001a,Schmid2009a} and superdense coding~\cite{Barreiro2008}. Hyperentanglement also has applications in optical tests of nonlocality~\cite{Chen2003}, as well as linear-optical quantum computing~\cite{Chen2007c,Vallone2010a} and the generation of multi-qubit entangled states using a fewer number of photons~\cite{Gao2010}. In this context, light-matter hyperentanglement was  demonstrated using spatial and polarization DOF, and was used in a demonstration of one-way quantum computing~\cite{Xu2012a}. 
The optical implementation of entanglement purification can be simplified greatly using hyperentanglement~\cite{Simon2002}. This could play an important role in the context of long-distance quantum communication with quantum repeaters, where purification can be used to increase the rate at which entanglement is distributed~\cite{Briegel1998,Duan2001}. However, this is possible only if the DOF in which the hyperentanglement is coded are suitable for long-distance transmission, e.g.~in optical fibre. Previous demonstrations of entanglement purification were all based on polarization and spatial modes~\cite{Pan2003,Walther2005a}, but the latter is not adequate for long-distance transmission in fibre. 
Energy-time (or time-bin) and polarization hyperentanglement is much better suited for this. 
The requirements that then arise for quantum repeaters is to have quantum memories that can efficiently store both DOF, combined with the possibility of efficiently distributing entanglement over long distances in optical fibre. 

Here we report on the quantum storage of hyperentanglement that is compatible for long-distance quantum communication in optical fibre. A source first generates photons hyperentangled in polarization and energy-time. One photon from the pair is then stored in a quantum memory based on rare-earth-ion doped crystals that is designed to store both DOF. The other photon has a telecommunication wavelength and can be distributed over long distances. 

The paper is organized as follows. In section~\ref{sec:ExpSet} we describe our experimental setup, including the source of hyperentangled photons. Details on the quantum memory are given in section~\ref{sec:qm}. Section~\ref{sec:chsh} describes how the Bell-CHSH inequalities on each DOF are measured, and section~\ref{sec:res} presents the main results.


\section{Experimental setup}
\label{sec:ExpSet}

\begin{figure*}[!t]
\centering
\def\svgwidth{1\textwidth}
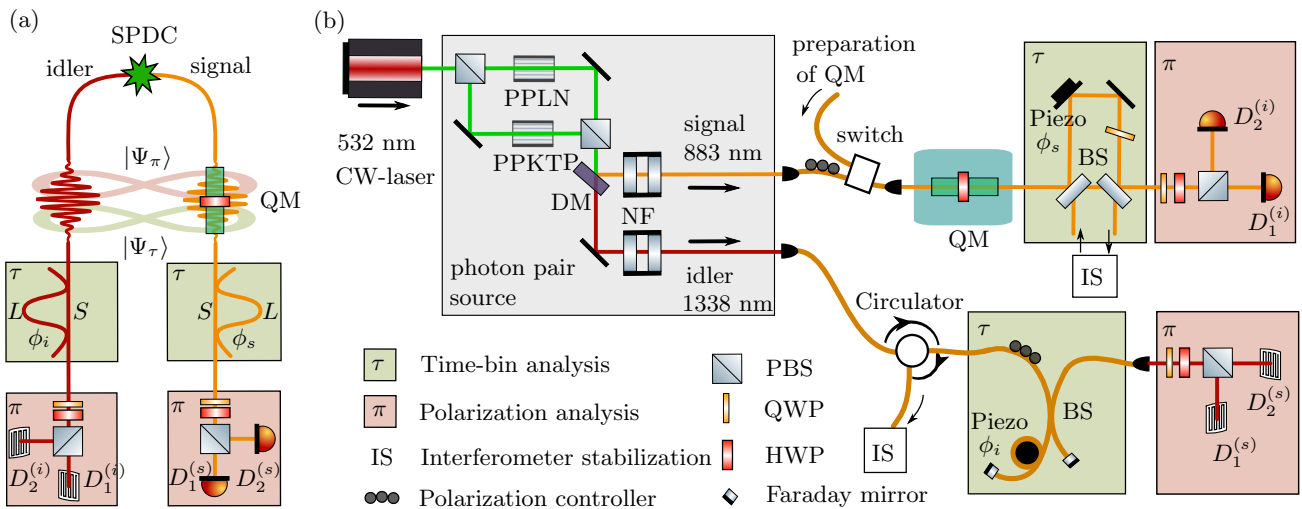
\caption{Experimental setup. (a) Conceptual setup of hyperentanglement storage inside a solid-state quantum memory (QM). A pair of photons entangled in polarization ($\ket{\Psi_{\pi}}$) and energy-time ($\ket{\Psi_{\tau}}$) are generated from SPDC. The signal photon is stored inside a quantum memory and released after a predetermined time of 50~ns. The hyperentanglement is revealed using time-bin analyzers ($\tau$) having short (S) and long (L) arms and adjustable relative phases ($\phi_i$ and $\phi_s$), followed and polarization analyzers $\pi$. (b) Experimental setup (see text for details). Polarization-entangled photon pairs are created by coherently pumping two nonlinear waveguides (PPLN and PPKTP) and recombining the optical paths. The pump is a CW laser at 532~nm, which inherently produces pairs that are also energy-time entangled. A dichroic mirror (DM) separates signal and idler photons. The appropriate linewidth for storage of the signal photon in the QM is obtained with narrow filtering (NF), which consists of a cavity and a volume Bragg grating for each the signal (883~nm) and the idler (1338~nm). An optical switch is used to direct either the light necessary for the preparation of the QM, or the signal photons, to the QM. The time-bin analyzers of the signal and idler photons are made with free-space and fiber components, respectively, using 50/50 beamsplitters (BS) and are both actively locked. Piezo elements are used to control the phases $\phi_s$ and $\phi_i$ of the analysers. They are followed by free-space polarization analyzers composed of quarter-wave and half-wave plates (QWP and HWP) followed by polarizing beam splitters (PBS). $D_{1,2}^{(s)}$  are avalanche photodiodes and  $D_{1,2}^{(i)}$ are WSi superconducting nanowire single-photon detectors.
}
\label{fig:exp}
\end{figure*}

A conceptual setup of our experiment setup is depicted in Fig.~\ref{fig:exp}(a). It consists of a source of pairs of entangled photons (denoted as signal and idler photons) entangled in both polarization and energy-time, a solid-state quantum memory based on rare-earth-ion doped crystals, analyzers (denoted as $\tau$) used to reveal the energy-time entanglement, followed by analyzers (denoted as $\pi$) used to reveal the polarization entanglement. 

Fig.~\ref{fig:exp}(b) shows a detailed version of our setup. Photon pairs entangled in both DOF (hyperentangled photon pairs), consisting of a signal photon at 883~nm and idler photon at 1338~nm, are produced by spontaneous parametric downconversion (SPDC) in nonlinear waveguides. Energy-time entanglement is obtained by pumping the waveguides with a continuous-wave laser at 532~nm with an average power of 2.5~mW. Photons from a given pair are created simultaneously at a time that is uncertain within the coherence time of the pump, which creates the entanglement. The polarization entanglement is generated by sending diagonally-polarized pump light onto a polarizing beam splitter (PBS) that transmits horizontal polarization and reflects vertical. The horizontal output of the PBS is followed by a periodically-poled lithium niobate waveguide (PPLN) oriented to ensure non-degenerate, collinear type-0 phase-matching generating horizontally polarized photons. Similarly, the vertical output is sent to a potassium titanyl phosphate waveguide (PPKTP) generating vertically polarized photons. After each waveguide, the signal and idler are separated by a dichroic mirror (DM) and the modes are recombined at a PBS. A photon pair is then in a coherent superposition of being emitted by the first ($\ket{HH}$) and second waveguide ($\ket{VV}$), yielding a state close to the maximally entangled Bell-state
\begin{equation}
\ket{\Psi_{\pi}}=\frac{1}{\sqrt{2}}(\ket{HH}+e^{i\theta}\ket{VV}). \label{bellp}
\end{equation}
Using the residual pump light collected at the unused output port of the PBS, we derive a feedback signal that is used to stabilize the phase $\theta$. The polarization entanglement of this source was described in detail in Ref.~\cite{Clausen2014}.

Storage of the photon in an atomic ensemble requires reducing its spectral width from its initial $\approx 500$~GHz linewidth down to a fraction of the storage bandwidth of the memory, which is $\sim 600$~MHz. The narrow filtering (NF) for the signal and idler photons is done in two steps: in each path, we combine a filtering cavity and a volume Bragg gratings (VBG) to select only a single longitudinal mode of the cavity. The idler photon first passes through a Fabry-Perot cavity with linewidth of 240~MHz and free spectral range (FSR) of 60~GHz. It is then followed by a VBG with a FWHM diffraction window of 27~GHz. The signal photon is first sent onto a VBG with a spectral bandwidth of 54~GHz and then sent through an air-spaced Fabry-Perot etalon with a linewidth of 600~MHz and FSR of 50~GHz. Due to the strong energy correlation between both photons, the heralded signal photon's linewidth is effectively filtered to $\approx 170$~MHz, corresponding to a coherence time $\tau_c \approx 1.9$~ns~\cite{Clausen2014}.

The signal photon is then sent for storage in a compact, polarization-preserving and multimode solid-state quantum memory (QM), as described in section~\ref{sec:qm}. It is then retrieved from the QM after a pre-determined 50~ns storage time with an efficiency of $\approx 5$\%.

To reveal energy-time entanglement of the photon retrieved from the memory a Franson interferometer~\cite{Franson1989} is used. Specifically, each photon is then sent through unbalanced interferometers with controllable phases and identical travel-time difference between the short (S) and long (L) arms (these interferometers are shown are all-fibre Mach-Zehnder interferometers on Fig.~\ref{fig:exp}(a)). In practice (Fig.~\ref{fig:exp}(b)), the idler is sent through an unbalanced all-fiber Michelson interferometer using Faraday mirrors, and the signal is sent through a free-space Mach-Zehnder interferometer. The travel-time difference between the short and long arms is 5.5~ns, which is greater than the coherence time of the photons $\tau_c$ and eliminates single-photon interference. However, due to the large uncertainty in the creation time, a coincidence stemming from both photons travelling the short arms is indistinguishable from one where both photons travelling the long arms, leading to quantum interference in the coincidence rate. 

Hence, interference fringes can be observed by varying the phase in each interferometer. These coincidences can be seen as stemming from a time-bin entangled state that is close to the maximally entangled Bell state 
\begin{equation}
\ket{\Psi_{\tau}} = \frac{1}{\sqrt{2}}(\ket{SS}+\ket{LL})\label{bellt}.
\end{equation}
Coincidences between photons travelling different arms are also observed, but they do not yield any kind of interference and are discarded when analyzing the energy-time DOF. They can however be kept when analyzing the polarization DOF.

When considering entanglement in both DOF, the state of a single pair can be written as 
\begin{equation}
\ket{\Psi_{\tau}}\otimes \ket{\Psi_{\pi}}. \label{eq:hypent}
\end{equation}
Both DOF of an hyperentangled pair can in principle be manipulated independently, and the quality of the entanglement in one DOF should not depend on the basis in which the other is measured. In our setup, this is possible only if polarization rotations, due to birefringent optics, are the same in both arms of the unbalanced interferometers. For the idler photon, this is happening automatically, thanks to the Faraday mirrors reflecting the light with a polarization that is orthogonal to the one at the input of the 50/50 fibre beam splitter (BS). 
For the signal photon, this is more challenging because free-space mirrors affect the phase of the $\ket{H}$ and $\ket{V}$ polarizations in different ways.
This effectively means that without any kind of compensation the measurement basis of the time-bin analyzer is not the same for an input $\ket{H}$ or an input $\ket{V}$ polarization state. To eliminate this problem, we insert a wave plate in the long arm. The fast axis is set to horizontal and the plate is tilted with respect the beam (see Fig.~\ref{fig:exp}). The tilt controls the relative phase between horizontal and vertical polarizations, and is adjusted to equalize the birefringence of both arms and therefore eliminate the polarization-dependent relative phase between the two arms.  

To lock the phase of the idler's time-bin analyzer, we use highly coherent light at 1338~nm obtained from difference-frequency generation (DFG) from 532 and 883~nm light combined in the PPLN waveguide. A feedback mechanism locks this DFG light on the idler's cavity transmission peak~\cite{Clausen2014}. The phase of the interferometer is controlled by coiling the fiber of the long arm around a cylindrical piezo transducer, and the interferometer is locked using a side-of-fringe technique. 
The phase of the signal photon's time-bin analyzer is controlled using a piezo-mounted mirror placed in the long arm. 
The phase is probed using part of the CW laser at 883~nm that is used to prepare the QM. The light is frequency shifted using an acousto-optic modulator (AOM) and then sent trough the interferometer in a spatial mode that has no overlap with the signal photon. 
The phase of the interferometer is modulated with a sinusoidal signal oscillating at 18~kHz. This yields intensity fluctuations that can be demodulated using a lock-in technique and allows us to obtain the derivative of the transmission of the interferometer.
To scan the phase, we fix the locking point to a maxima of the transmission (i.e.~a zero of its derivative) and sweep the frequency of the probe laser using the AOM. The time difference of 5.5~ns between the short and long arms yield a period that can be covered by scanning the frequency over $\approx180$~MHz. One advantage of this technique compared to a side-of-fringe lock is that it yields a locking point that is unaffected by fluctuations of the intensity of the laser that probes the phase.

After the interferometers, the polarization of each photon is analyzed. Each output of the PBS is coupled into a single-mode fiber and sent to single-photon detectors. The results of the measurements made at different analyzers are compared in order to reveal the nonlocal correlations in both DOF. Single-photon detectors with 30\% (Si avalanche photodiode) and 75\% (WSi superconducting nanowire~\cite{Verma2014a}) efficiencies are used to detect signal at 883~nm and idler at 1338~nm, respectively.

The heralding efficiency of signal photons up to the quantum memory is $\approx 20\%$, while the overall detection efficiency of idler photons is $\approx 10\%$. The average input pump power at 532~nm of 2.5~mW was used and corresponds to a photon pair creation probability of $\approx 0.015$ for the time window of $\tau_c = 1.9$~ns~\cite{Clausen2014}. The overall coincidence rates for the transmitted and stored photons were $\approx 20$~Hz and 2~Hz, respectively.

\section{Multimode and polarization-preserving broadband Quantum Memory}
\label{sec:qm}

In this section we describe our quantum memory and how it can store both DOF. 
The storage is implemented using the Atomic Frequency Comb (AFC) storage protocol in rare-earth-ion doped crystals~\cite{Afzelius2009a}. To realize this, the inhomogeneously broadened absorption profile of the crystal is first shaped into a comb-like structure in frequency using optical pumping. When a photon is absorbed by the AFC, it creates an atomic excitation delocalized over all atoms inside the comb. The collective state then dephases and the excitation is stored. Thanks to the periodic profile of the AFC, the atoms then collectively interfere after a specific time, which can lead to re-emission of the signal photon into the same spatial mode it was absorbed in. 
The storage time is pre-determined and equal to $1 / \Delta$, where $\Delta$ is the period of the frequency comb.

The temporal multimode capacity for this protocol is given by the ratio of the storage time over the duration of the temporal modes that are stored. For a given storage time, the multimode capacity therefore increases with the storage bandwidth. The large inhomogeneous broadening of rare-earth-ion doped crystals makes them an excellent material to realize multimode quantum memories at the single-photon level~\cite{Usmani2010}, and are well suited for the storage of energy-time and time-bin entanglement, as demonstrated in~\cite{Clausen2011,Saglamyurek2011}. 

\begin{figure}[!t]
\centering
\def\svgwidth{0.5\textwidth}
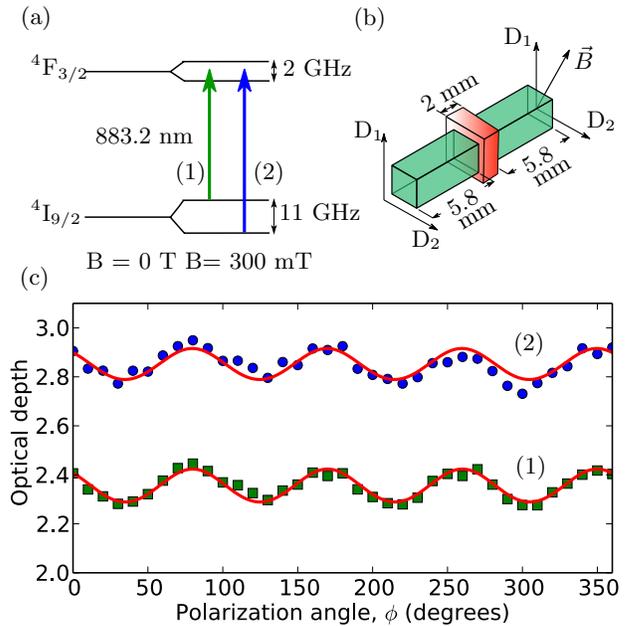
\caption{Scheme for storage of polarization qubits. (a) Energy level structure of Nd$^{3+}$ ions inside a \YSO{} crystal with and without applied external  magnetic field $B=300$~mT. External magnetic field lies in  $D_1$-$D_2$ plane with $30^{\circ}$ angle with respect to $D_1$ axis;
(b) The compact configuration of the quantum memory is obtained by placing a HWP between two identical \nd{} crystals. The fast axis of the HWP is oriented at 45$^{\circ}$ with respect to the axes $D_1$ and $D_2$, which are the two of the principal axes of the dielectric tensor. The 14~mm long arrangement is cooled to 2.7~K and placed in a static magnetic field to split the ground in two Zeeman levels. 
(c) The optical depth of the two-crystal configuration is shown as a function of the linear polarization angle of the input. The green squares and blue circles correspond to transitions ((1) and (2)) from each of the Zeeman-split ground states shown in (a). Lines are fits of the model described in~\cite{Afzelius2010c}.}
\label{fig:od}
\end{figure}

We implement the AFC quantum memory protocol using rare-earth ion doped \nd{} crystals with a dopant concentration of $\approx 75$~ppm. Optical pumping is used to shape the absorption profile of the QM in an atomic frequency comb. This requires splitting the ground state $^4I_{9/2}$ in two Zeeman levels~(Fig.~\ref{fig:od}(a)) using a static magnetic field of 300~mT~\cite{Usmani2010}. This is done to spectrally resolve two optical transitions which are inhomogeneously broadened to 6~GHz and to perform optical pumping from one ground Zeeman state to another. The Zeeman splitting of the excited state is not spectrally resolved in this configuration. We measured a ground state Zeeman lifetime of $43$~ms using spectral hole burning measurements, which is much greater than the 300~\textmu s radiative lifetime of the optical transition, as required for optical pumping. We note this lifetime is however shorter than the $\sim100$~ms measured in 30~ppm-doped crystals~\cite{Usmani2010}, which unavoidably affects the quality of the AFC that we can prepare; see below.

The efficiency $\eta$ of the AFC protocol~\cite{Afzelius2009a} depends on the optical depth $d$ through $\eta = \tilde{d}^2 \e^{-\tilde{d}}\e^{-d_0}\eta_{\text{deph}}$, where $\tilde{d}=\frac{d}{F}$ is the average optical depth of the comb, $F$ is the finesse of the comb, $d_0$ the residual optical depth and $\eta_{\text{deph}}$ is the dephasing term, which is maximized for square peak shapes~\cite{Bonarota2010}. The storage of polarization qubits in rare-earth-ion doped crystals is therefore hindered by their polarization-dependent optical depth. It is however possible to mitigate this problem, as demonstrated in~\cite{Clausen2012,Gundogan2012,Zhou2012}. 
Specifically, consider a crystal cut such that its input face contains two principal axes of the dielectric tensor; see Fig.~\ref{fig:od}(a). Let $D_1$ and $D_2$ be those axes, which we assume coincident with the polarizations for which the optical depth is minimum and maximum. This condition is satisfied for an yttrium orthosilicate crystal doped with neodymium ions~\cite{Clausen2012}, \nd{}, the material we use here. Let $d_1$ and $d_2$ be the optical depth for light polarized along $D_1$ and $D_2$ axes. By placing two identical crystals on each sides of a half-wave plate (HWP) oriented to rotate a $D_1$-polarized photon to $D_2$ and vice versa, an absorbed single photon with an arbitrary polarization will be in superposition of being stored in both crystals with an effective optical depth equal to $d_1 + d_2$, yielding a polarization-independent efficiency.

\begin{figure}[!t]
\centering
\def\svgwidth{0.45\textwidth}
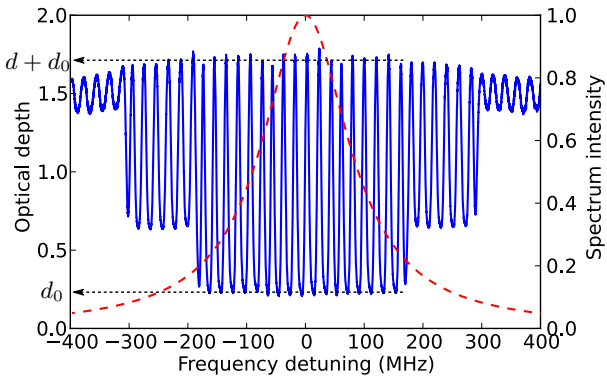
\caption{Spectrum of the AFC prepared by optical pumping inside the absorption profile. The central 120~MHz-wide region is prepared by the carrier frequency of the laser diode that is modulated in intensity and frequency by an acousto-optic modulator. The subsequent 120~MHz-wide regions on both sides are prepared by generating first and second-order sidebands separated by 120 and 240~MHz from the carrier frequency, respectively, using an electro-optic phase modulator placed after the AOM. The finesse of the comb is $\approx2$ and the width of the comb is $\approx 600$~MHz. For comparison, the dashed red line shows the power spectra of a 170~MHz Lorentzian, which is close to the spectral width of the heralded signal photon.  The values of $d$ and $d_0$ used in equation of efficiency are shown for the central part.
}
\label{fig:afc}
\end{figure}

Here we realize this polarization-independent scheme in a compact manner using two 5.8 mm-long \nd ~crystals placed on each sides of a 2 mm-thick HWP, resulting in a total length of about 14~mm (Fig.~\ref{fig:od}(b)). This setup system is cooled to a temperature of 2.7~K using a closed-cycle cryocooler. The Nd$^{3+}$ dopant concentration is higher than we used in a previous demonstration of polarization-independent storage of single photons~\cite{Clausen2012}. The higher doping level allows us to use shorter crystals for more compact setup.  All faces of the crystals, HWP and cryostat windows are coated with anti-reflective films and the overall transmission through the system is higher than 95\%, when factoring out the absorption of the crystals. 

We achieve an average optical depth of $2.35 \pm 0.10$, averaged over all linear polarization states, for the optical transition starting from the  higher-energy level of the Zeeman doublet.  
The variations are smaller than 5~\%~(Fig.~\ref{fig:od}(c)). They may be attributed to the imperfect alignment and retardation of the HWP, as well as an imperfect optical alignment of the beam with respect to the input face. 
The transition from the other Zeeman level yields $2.85\pm 0.11$ (see Fig.~\ref{fig:od}(c), blue circles). This difference is consistent with thermalized populations, which dictates that the ratio of the optical depths should be $\exp(-\Delta E/k_B T)$, where $\Delta E = h\Delta \nu$, $\Delta \nu = 11$~GHz, $k_B$ is Boltzmann's constant and $T$ is the temperature. For our temperature of 2.7~K, the expected ratio is 0.82, which matches the observed ratio of $2.35/2.85 = 0.825$.  For our storage of hyperentangled photons, we could only use the optical transition with the lowest optical depth due to the limited tuning range of the Fabry-Perot etalon of the source. Using the other optical transition could have led to a higher quantum memory efficiency.

The AFC is prepared with an AOM that modulates the intensity and frequency of the light from an external cavity diode laser centred on the absorption line of the $^4I_{9/2}\rightarrow~ ^4F_{3/2}$ transition. In this way we create a 120~MHz absorption comb with a spacing of $\Delta=20$~MHz between the peaks, which corresponds to $1/\Delta=50$~ns storage time~\cite{Afzelius2009a}. To extend the bandwidth of QM beyond the 120~MHz limit imposed by the double-pass in the AOM, the light from the AOM is coupled inside an electro-optical phase modulator that creates first and second order sidebands separated by 120~MHz from each other. 
This yields an overall comb width of 600~MHz, larger than the 170~MHz spectrum of the heralded signal photon, as shown in Fig.~\ref{fig:afc}. The contrast of the absorption profile is reduced on the sides due to the smaller optical power available in the second-order sidebands. 
The fact that the maximum optical depth of the comb (1.8) is less than the one of the transition itself (2.35) can be attributed to power broadening, which can reduce the maximum optical depth between the peaks. The remaining absorption of $d_0\ge 0.25$ also reduces the storage efficiency. The efficiency of the quantum memory with this comb is $\approx 5$\% for a 50~ns storage time, while the total absorption and the transmission of the memory both are close to 50\%. Imperfect rephasing and reabsorption  processes inside the memory lead to the decrease of the QM efficiency~\cite{Afzelius2009a}. The photons that were not absorbed could be used to analyze storage process and calibrate the analyzers for CHSH inequality violation.

\section{Bell tests on hyperentanglement}
\label{sec:chsh}

A quantum state is hyperentangled if one can certify entanglement for each entangled DOF. 
Therefore, it is enough to violate a Bell inequality in both polarization and time independently to demonstrate hyperentanglement. Here we use the inequality derived by Clauser et al.~\cite{clauser1969} (CHSH) to witness entanglement. The Bell-CHSH inequality for a single DOF reads
\begin{eqnarray}
S &=& \lvert E(\X, \Y) + E(\X^{'}, \Y) + E(\X, \Y^{'}) - E(\X^{'}, \Y^{'})\rvert\nonumber\\ &\leq& 2\label{eq:chsh}
\end{eqnarray}
where $\X$ and $\X^{'}$ ($\Y$ and $\Y^{'}$) are two observables that are measured on the signal (idler) side. $E$ is the correlator corresponding to the expectation value of the correlation between measurement results obtained on both photons of an entangled pair. Those correlators are calculated from the number of coincidences between idler detector $D_k^{(i)}$ and signal detector $D_l^{(s)}$, denoted $R_{kl}$, where $k,l \in \{1,2\}$ are the possible outcomes for each measurement. In terms of the coincidence rate, the correlation function is written as $E = [R_{11}+R_{22}- R_{12}- R_{21}]/R_{T}$, where $R_T = \sum_{k,l}R_{kl}$ is the total rate of coincidences. 

Let us first consider the polarization DOF only, and suppose for now that the time-bin analyzers in the setup of Fig.~\ref{fig:exp} are bypassed. Consider as well that the signal and idler photons are measured in the set of linear polarizations. This is done by setting the QWPs of the analyzers at 0$^{\circ}$ with respect to horizontal, and the HWPs at $\theta_s$ and $\theta_i$ for the signal and idler, respectively. Finally, let us assume the phase $\theta$ of state $\ket{\Psi_{\pi}}$ is equal to zero (Eq.~\ref{bellp}).  One can show that the coincidence rates should be $R^{(\pi)}_{11}=R^{(\pi)}_{22} \sim (1+V_{\pi}\cos \left[4(\theta_s - \theta_i)\right])$, and $R^{(\pi)}_{12}=R^{(\pi)}_{21} \sim (1-V_{\pi}\cos \left[4(\theta_s - \theta_i)\right])$. In these expressions we introduced the polarization entanglement visibility $0\le V_{\pi}\le 1$ that arises by assuming the experimental imperfections can be described by replacing $\ket{\Psi_{\pi}}$ by a Werner state of visibility $V_{\pi}$~\cite{Werner1989a}.

\begin{figure}[!t]
\centering
\def\svgwidth{0.45\textwidth}
\input{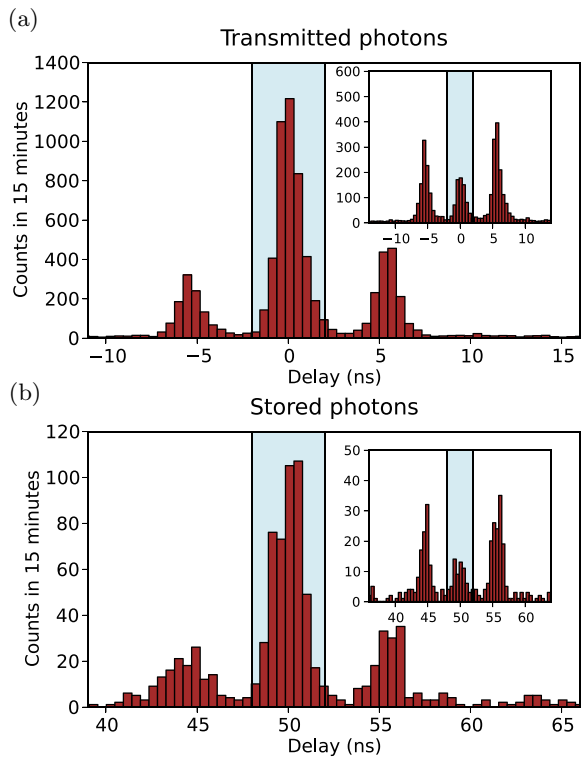}
\caption{Example of measurements used to violate CHSH inequality for time-bin qubits. The coincidence histograms between detectors $D_{1}^{(s)}$ and $D_{1}^{(i)}$ shows three peaks corresponding to different path combinations for (a) a transmitted signal photon, i.e.~not absorbed by the QM, and (b) a stored signal photon. 
Each figure represents an histogram from one measurement outcome, $R_{11}^{(\pi,\tau)}$, of a correlator in the Bell-CHSH inequality (Eq.~\ref{eq:chsh}). The insets correspond to histograms with an additional $\pi$~phase shift between the two interferometers. Varying the angles of the polarization analyzers lead to variations of the intensity of all three peaks simultaneously.}
\label{fig:delay}
\end{figure}

Let us now consider the case where the interferometers are inserted before the polarization analyzers, as in Fig.~\ref{fig:exp}. This yields a Franson interferometer, and one can show that the observed coincidence rate between $D^{(s)}_{k}$ and $D^{(i)}_{l}$ (when considering the appropriate time difference between detections; see below) is given by 
\begin{equation}
R^{(\pi,\tau)}_{kl} = R^{(\pi)}_{kl}\cdot R^{(\tau)},
\end{equation}
where $R^{(\tau)} \sim (1+V_{\tau}\cos \left[\Delta \phi_s + \Delta \phi_i\right])$, and $\Delta \phi_s$ (or $\Delta \phi_i$) is the relative phase between the long and short arms of the signal (or idler) interferometer. Like the polarization DOF, we assumed the measurement is performed on a Werner state of visibility $V_{\tau}$ instead of $\ket{\Psi_{\tau}}$. Note that in our setup, we use only one of the output ports of the interferometers. This translates into saying that $R^{(\tau)}$ corresponds to one of the four rates possible $R^{(\tau)}_{mn}$ ($m,n \in \{1,2\})$ at a time. Which one is measured is decided by an appropriate choice of the relative phases $\Delta \phi_{s,i}$. This limitation can nevertheless be used to violate a Bell-CHSH inequality, assuming fair sampling of the outcomes. 

When measuring the Bell-CHSH inequality on the polarization DOF only, $R^{(\pi)}_{kl}$ is obtained from a measurement of $R^{(\pi,\tau)}_{kl}$ by considering $R^{(\tau)}$ as a constant loss factor. Measurement of $R^{(\tau)}$ is obtained by summing all coincidences between either $D^{(s)}_1$ or $D^{(s)}_2$ and $D^{(i)}_1$ or $D^{(i)}_2$. Quantum mechanics predicts that $S_{\pi} \le 2\sqrt{2} V_{\pi}$ (or $S_{\tau} \le 2\sqrt{2} V_{\tau}$) for polarization (or energy-time), where the inequality is saturated with an optimal set of measurements. The local bound is $S_{\pi,\tau} =2$.

\section{Results}
\label{sec:res}

Before characterizing quantum correlations, we first need to determine the phase $\theta$ of the polarization-entangled state $\ket{\Psi_{\pi}}$, as well as the sum $\Delta \phi_s + \Delta \phi_i$ of the phases of the interferometers. Once they are known, we consider them as phase offsets in $R^{(\pi)}_{kl}$ and $R^{(\tau)}$. For this, we use the signal photons that are transmitted by the quantum memory (i.e.~not stored). We do so because the signal photons are more likely to be transmitted than stored and retrieved, and this allows a faster characterization of the phases. Figure~\ref{fig:delay}(a) shows coincidence histograms between the pair of detectors $D_{1}^{(s)} \mbox{and} \,D_{1}^{(i)}$ (the other histograms are similar). The coincidences are resolved into two satellite peaks and a central peak. The rate estimated from the central peak corresponds to $R_{11}^{(\pi,\tau)}$. The rate in the satellite peaks is proportional to $R_{11}^{(\pi)}$ only since the timing between detections cannot lead to quantum interference due to the energy-time entanglement. Hence, the satellite peaks are included in the analysis of the polarization entanglement, but not in the one of the energy-time.

\begin{figure}[!t]
\centering
\def\svgwidth{0.45\textwidth}
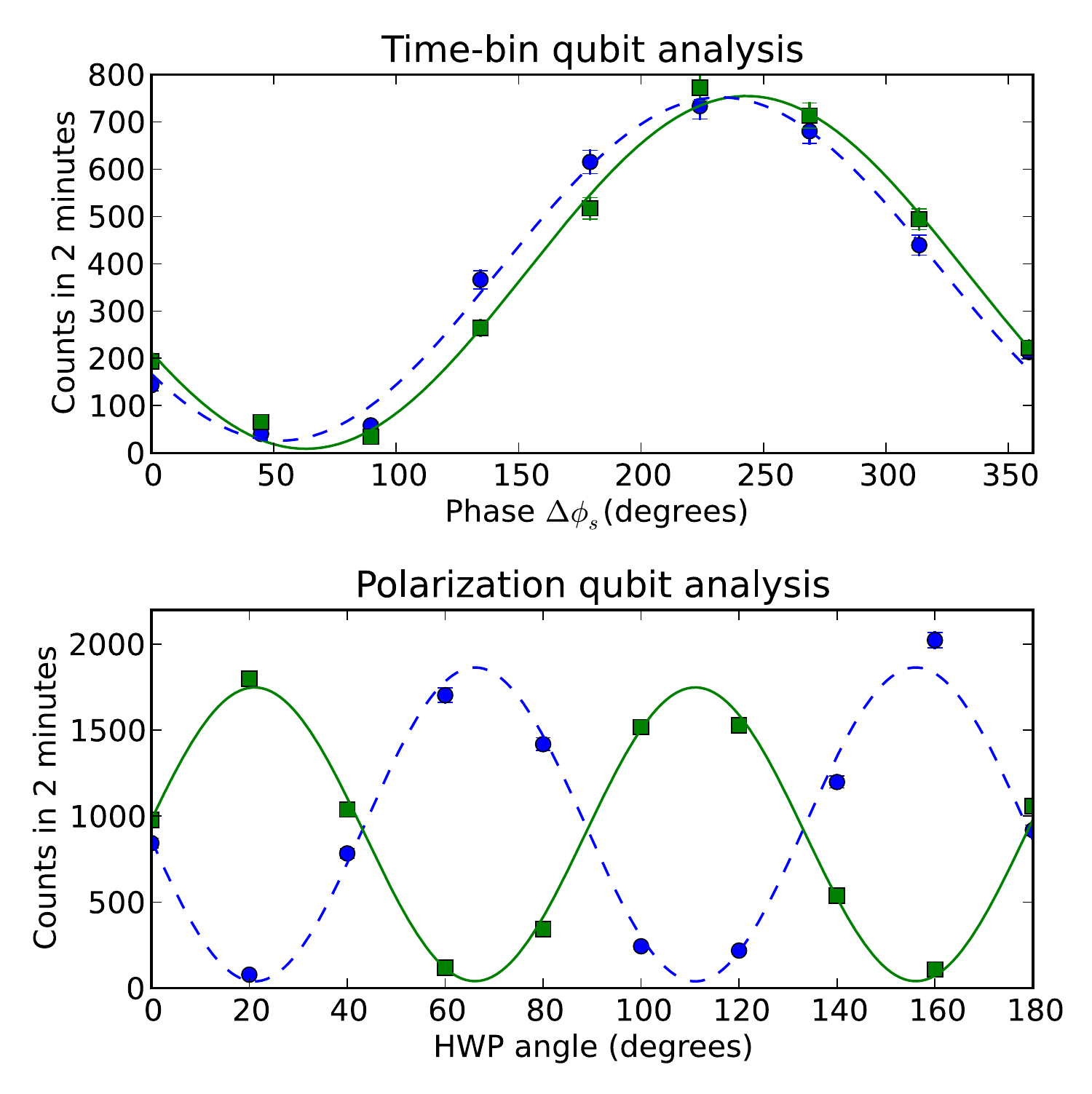
\caption{Calibration of the analyzers using transmitted signal photons. (a) Rates in the central coincidence peak of Fig.~\ref{fig:delay}(a) are plotted as a function of the sum of the phases of each interferometers, for both output ports of the polarization analyzer. The small phase shift between the curves appears due to a residual phase difference between $\ket{H}$ and $\ket{V}$ components at the output of the interferometer on the signal side. (b) Rates in the central coincidence peak of Fig.~\ref{fig:delay}(a) as a function of polarization analyzer's HWP angle of the signal photon (with the QWP at $45^{\circ}$), for two pairs of detectors, namely $D_1^{(s)} \mbox{and } D_1^{(i)}$ (solid line) or $D_2^{(s)} \mbox{and } D_2^{(i)}$ (dashed line). Each curve represents a fit to data points using a sinusoidal function and the error bars are estimated assuming Poisson statistics for the counts. The average visibilities for polarization and energy-time entanglement are $V_\tau=92(3)~\%$ and $V_\pi=96(2)~\%$, respectively.}
\label{fig:pol_time}
\end{figure}

To extract $\Delta \phi_s + \Delta \phi_i$, we scan the free-space interferometer as described in section~\ref{sec:ExpSet}, while the polarization is measured in the basis $\{\ket{H},\ket{V}\}$ on both sides. In this way, detectors $D_1^{(s)}$ and  $D_1^{(i)}$ are revealing the energy-time entanglement of the $\ket{HH}$ component of $\ket{\Psi_{\pi}}$, while $D_2^{(s)}$ and  $D_2^{(i)}$ are revealing the one of the $\ket{VV}$ component. 
Fig.~\ref{fig:pol_time}(a) shows coincident events in the central peak in function of $\Delta\phi_{s}$, while  $\Delta\phi_{i}$ is kept constant. The rates $R_{11}^{(\pi,\tau)}$ (solid line) and $R_{22}^{(\pi,\tau)}$ (dashed line) overlap, as it is required to measure both DOF independently. The visibility is $V_{\tau} = 92(3)\%$. 
To extract the value of $\theta$, we set $\Delta \phi_s + \Delta \phi_i = 0$ and project the idler photon in the basis $\{\ket{+},\ket{-}\}$, where $\ket{\pm}=\frac{1}{\sqrt{2}}(\ket{H}\pm\ket{V})$. This prepares the signal photon in the state $\frac{1}{\sqrt{2}}(\ket{H}\pm e^{i\theta}\ket{V})$. The QWP of the signal polarization analyzer is then set to transform that state into one with a linear polarization. Fig.~\ref{fig:pol_time} (b) shows the rates $R_{11}^{(\pi)}$ and $R_{22}^{(\pi)}$, as a function of the HWP angle, from which the phase of the polarization Bell-state can be extracted from the horizontal offset. The visibility $V_{\pi} = 96(2)\%$.

\begin{figure}
\centering
\def\svgwidth{0.59\textwidth}
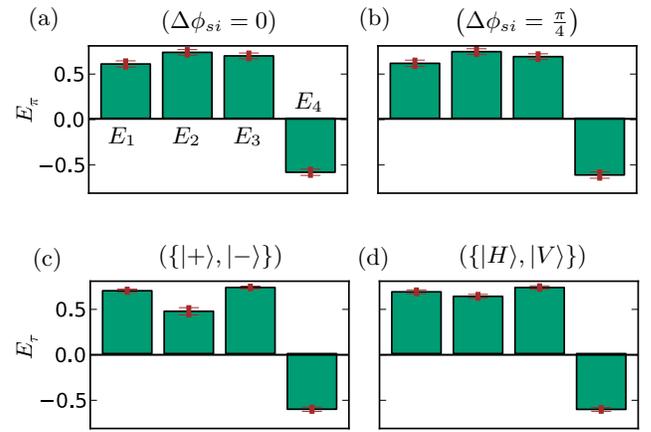
\caption{Correlators for stored photons. The four panels are different sets of correlation measurements that violate the Bell-CHSH inequality of \eqref{eq:chsh} reported in Table~\ref{tbl:sum}. The top row shows the values for polarization measurements with either (a) measurement on the energy-time entanglement such that $\Delta\phi_{s}+\Delta \phi_i=0$, or (b) with $\Delta\phi_{s}+\Delta \phi_i=\pi/4$. In the bottom row we show the values for energy-time measurements while projecting the polarization of both the signal and idler in (c) the $ \left\{\ket{+},\ket{-}\right\}$ basis, or (d) the $ \left\{\ket{H},\ket{V}\right\}$ basis (right). }
\label{fig:chsh}
\end{figure}

Once the phases are estimated, we measure correlations that violate the Bell-CHSH inequality for each DOF using the photons that are stored and retrieved from the quantum memory. To illustrate the independence between the two DOF, each polarization measurement was performed using two different projection bases for the energy-time DOF, and vice versa. Specifically, the test on the energy-time DOF was done with using either the polarization basis $\pi_1 = \left\{\ket{H},\ket{V}\right\}$ for both photons, or $\pi_2 = \left\{\ket{+},\ket{-}\right\}$. The test on the polarization DOF was done with either $\Delta\phi_{s}+\Delta \phi_i = \phi_{si} = 0$ (denoted by $\tau_1$) or $\phi_{si} = \frac{\pi}{4}$ (denoted by $\tau_2$). 
The values of the measured correlators are shown on Fig.~\ref{fig:chsh}. 
The corresponding CHSH parameters are $S_{\pi}^{(\tau_1)} = 2.59(4)$ and $S_{\pi}^{(\tau_2)} = 2.64(4)$ for polarization entanglement, and $S_{\tau}^{(\pi_1)} = 2.60(7)$ and $S_{\tau}^{(\pi_2)} = 2.49(4)$ for the energy-time. The violations exceed the local bound by more than 8~standard deviations. 
To see the effect of the storage process on the quality of the hyperentanglement, we performed the same analysis using the transmitted signal photons. Table~\ref{tbl:sum} summarizes all the results. 

The values for $S_{\pi}$ and $S_{\tau}$ for stored photons are all lower than for transmitted photons except the $S_{\tau}$ value for $\pi_1$ polarization basis (which we believe is higher due to a statistical fluctuation). The lower values are most likely caused by accidental coincidences between idler photon (from one photon pair) and signal photon (from another photon pair) generated within the time delay equal to the storage time of the memory. This effect was studied in detail in a previous publication~\cite{Usmani2012}. It limits the maximally achievable visibility and reduces the CHSH inequality violation. The relative difference between the transmitted and the stored cases at most 5~\%, and hence the storage in the QM has very little effect on the quality of the hyperentanglement.

\begin{table}[htp]
\centering
\renewcommand{\arraystretch}{1.5}
\begin{tabular}{ccc|cc}
                                  & \multicolumn{2}{c}{$S$ (transmitted)}     & \multicolumn{2}{c}{$S$ (stored)} \\
                                   & $\pi$                  & $\tau$                  & $\pi$    & $\tau$    \\
\cline{2-5}
 $\pi_1 : \left\{ \ket{H},\ket{V} \right\}$              & -                  &                   2.555(13) &      -   & 2.60(7)   \\
 $\pi_2 : \left\{ \ket{+},\ket{-} \right\}$              & -                  & 									 2.571(11) &      -   & 2.49(4)   \\
 $\tau_1: \phi_{si} =0$             & 								2.716(11) &   -                 & 2.59(4)  &     -     \\
 $\tau_2: \phi_{si} =\frac{\pi}{4}$ &                 2.733(12)  &   -                 & 2.64(4)  &     -     \\
\end{tabular}
\caption{
Summary of all Bell-CHSH violations. Measured $S$ parameters obtained with transmitted or stored signal photons are shown.  For each case, the tests on energy-time ($\tau$) DOF were done with either the polarization basis $\pi_1=\left\{\ket{H},\ket{V}\right\}$ or $\pi_2=\left\{\ket{+},\ket{-}\right\}$, and tests on the polarization ($\pi$) DOF was done with $\Delta\phi_{s}+\Delta \phi_i = \phi_{si} = 0$ ($\tau_1$) or $\frac{\pi}{4}$ ($\tau_2$). These results show clear violations of Bell-CHSH inequality and demonstrate entanglement in all DOF studied.}
  \label{tbl:sum}
\end{table}

\section{Conclusions}
\label{sec:conc}
We have shown the storage of energy-time and polarization hyperentanglement in a solid-state quantum memory. This choice of degrees of freedom, combined with the fact that one of the photons of each pair is at a telecom wavelength, makes our source and memory very attractive for the implementation of linear-optical entanglement purification in quantum repeaters. This would ultimately require the use of a quantum memory that can retrieve photon on-demand using a complete AFC storage scheme~\cite{Afzelius2009a,Timoney2013}. Alternatively, a scheme based on spectral multiplexing~\cite{Sinclair2014} could be used. The storage of hyperentanglement demonstrated here is suitable with both of the approaches. Our experiment also shows that we can store two entanglement bits in a single quantum memory. Expanding on this idea, our memory could be used to store other degrees of freedom, and an even larger number of entanglement bits, by using frequency, orbital angular momentum and spatial modes. The multimode capacity of rare-earth-ion doped quantum memories goes beyond the temporal degree of freedom, and this might prove a useful tool for optical quantum information processing.

\section{Acknowledgements}

We thank Christoph Clausen, Anthony Martin and Nicolas Sangouard for useful discussions. 
 
\section*{Funding Information}
This work was financially supported by the Swiss National Centres of Competence in Research (NCCR) project Quantum Science Technology (QSIT) and the European project SIQS. J.~L. was supported by the Natural Sciences and Engineering Research Council of Canada (NSERC).

\bibliography{hyper-bib}


\end{document}